\documentclass[reprint,twocolumn,superscriptaddress,amsmath,amssymb,prb,nofootinbib]{revtex4-2}
\usepackage[dvipdfmx]{graphicx}
\usepackage{dcolumn}
\usepackage{bm}
\usepackage{color}
\usepackage{amssymb}
\usepackage{comment}

\usepackage{ulem}

\begin{document}

\title{Ultrafast Spin Injection in Graphene via Dynamical Carrier Filtering at Transition Metal Dichalcogenide Interfaces}

\author{Shunsuke Yamada}
\affiliation{Kansai Institute for Photon Science, National Institutes for Quantum Science
and Technology (QST), Kyoto 619-0215, Japan}
\author{Arqum Hashmi}
\affiliation{Department of Nuclear Engineering and Management, Graduate School of Engineering, The University of Tokyo, 7-3-1 Hongo, Bunkyo-ku, Tokyo 113-8656, Japan}
\author{Tomohito Otobe}
\affiliation{Kansai Institute for Photon Science, National Institutes for Quantum Science
and Technology (QST), Kyoto 619-0215, Japan}

\date{\today}

\begin{abstract}
We report a real-time first-principles study of ultrafast spin injection in a WSe$_2$-graphene heterobilayer under circularly polarized laser irradiation, using time-dependent density functional theory. 
Contrary to conventional expectations, spin transfer into graphene is not a passive process but is actively driven by spin-selective carrier filtering at the interface. 
Spin-polarized carriers generated in the WSe$_2$ layer induce a preferential migration of opposite-spin carriers from graphene, which results in net spin magnetization in graphene. 
This process is governed by interlayer band offsets, density-of-state asymmetry, and Pauli blocking.
These findings indicate a microscopic mechanism of spin injection in non-magnetic systems and identify a guiding principle for the design of ultrafast opto-spintronic functionalities in van der Waals heterostructures.
\end{abstract}
\maketitle

\section{Introduction}

Two-dimensional (2D) materials such as transition metal dichalcogenides (TMDs) and graphene have emerged as promising platforms for spintronic applications, where the spin degrees of freedom are utilized as information carriers~\cite{Feng2017,Choudhuri2019,Liu2020,Ahn2020}. 
In particular, graphene exhibits exceptional carrier mobility and long spin diffusion lengths that exceed several micrometers, which makes it ideal for spin transport~\cite{Dlubak2012,Han2014}.  
However, the intrinsically weak spin-orbit coupling (SOC) in graphene limits the efficient generation and manipulation of spin polarization, which poses a challenge for its integration into spintronic devices~\cite{Roche2015}.

To address this limitation, van der Waals heterostructures composed of graphene and TMDs have attracted considerable attention~\cite{Sierra2021}.  
TMD monolayers possess strong intrinsic SOC due to the presence of heavy metal atoms and exhibit valley-dependent spin polarization that arises from spin-valley locking~\cite{Xiao2012,Mak2012,Xu2014}.  
A first-principles computational study predicted ultrafast magnetization via SOC in non-magnetic 2D materials including TMD monolayers under circularly polarized laser pulses on the femtosecond timescale~\cite{Neufeld2023}.
Although graphene cannot generate spin magnetization optically due to its negligible SOC, its high mobility makes it an ideal medium for spin transport.  
Therefore, efficient spin transfer from a TMD layer to a graphene layer in a stacked heterostructure could enable opto-spintronic devices that combine efficient spin generation and long-range spin transport.

Such spin injection and transfer processes have been experimentally demonstrated in several TMD-graphene heterostructures using circularly polarized light pulses~\cite{Luo2017,Du2018,Lorchat2018,Zhou2021}.  
These systems have also exhibited several spintronic functionalities, such as spin-charge conversion via the spin Hall and Rashba-Edelstein effects~\cite{Safeer2019,Ghiasi2019}, and tunable spin relaxation dynamics~\cite{Omar2017,Ghiasi2017}.
In particular, spin injection in TMD-graphene heterostructures has been discussed phenomenologically; however, its microscopic mechanism remains largely unexplored.

Most theoretical studies on TMD-graphene heterostructures have focused on their static band structures~\cite{Gmitra2015,Gmitra2016,Gmitra2017,Yang2018,Li2019,David2019,Wakamura2019,Naimer2021,Wang2022}.  
In contrast, real-time first-principles calculations based on time-dependent density functional theory (TDDFT)~\cite{Runge1984} provide a robust framework for the simulation of ultrafast spin and charge dynamics beyond empirical models.  
Previous TDDFT studies on ultrafast spin dynamics in 2D heterostructures~\cite{He2021,He2023,Li2023,Guo2024} have primarily addressed the optical intersite spin transfer (OISTR) effect in magnetic materials~\cite{Elliott2016,Dewhurst2018}.  
A previous first-principles study of electron dynamics in TMD-graphene heterostructures addressed charge transfer, but it did not consider spin degrees of freedom~\cite{Iida2018}.  
The ultrafast dynamics of spin injection in TMD-graphene heterostructures via laser pulses remains unexplored to date. 

In this study, we perform real-time first-principles simulations of ultrafast spin and charge dynamics in a hetero-bilayer (HB) of WSe$_2$ and graphene using TDDFT.  
We investigate the time evolution of spin-polarized carrier excitation and its transport within the HB under circularly polarized laser irradiation.  
To reveal a dynamical mechanism of spin injection in TMD–graphene heterostructures, we conduct first-principles calculations under various conditions, focusing on the dependence of spin transfer between WSe$_2$ and graphene on laser intensity.  
In these calculations, we focus on the initial stage of spin injection in the femtosecond timescale and neglect relaxation processes.
Our work provides microscopic insights into the dynamical response of TMD–graphene heterostructures and paves the way for the design of ultrafast opto-spintronic devices.

The remainder of this paper is organized as follows: Sec. II describes the theoretical methods and numerical conditions. 
In Sec. III, the calculation results are presented and analyzed in detail. 
Finally, the conclusions are presented in Sec. IV.


\section{Computational method}

\subsection{Structure optimization\label{sec:structure}}

The lattice constants of monolayer WSe$_{2}$ and graphene, which are 3.32 {\AA} and 2.46 {\AA}, respectively, were used to construct the bilayer heterostructure. 
The HB was modeled using a 2\texttimes{}2 supercell for WSe$_{2}$ and a 3\texttimes{}3 supercell for graphene, with the graphene layer rotated by 19{\textdegree}.
This configuration leads to a lattice mismatch within a tolerable range of 1.65\%.
Structural relaxation is performed using the OpenMX code, which is based on the linear combination of pseudoatomic orbitals (LCPAO) formalism \citep{Ozaki2003, Ozaki2004}. The pseudoatomic basis sets used are s$^{2}$p$^{2}$d$^{1}$ for carbon and s$^{3}$p$^{2}$d$^{2}$ for both tungsten and selenium, with norm-conserving pseudopotentials taken from the OpenMX library\cite{VPS_web}. 
Exchange-correlation effects are treated with the local spin density approximation (LSDA), namely the LSDA-CA functional \citep{Perdew1992}.  
The Brillouin zone is sampled using a 15\texttimes{}15\texttimes{}1 k-point grid, and an energy cutoff of 320 Ry is employed. 
A vacuum layer of 20 {\AA} is introduced along the \textit{z}-direction to eliminate spurious interactions between periodic images. 
During ionic relaxation, the shape and size of the supercell are fixed, while allowing for full relaxation of the atomic positions.
The optimization proceeds until the Hellmann–Feynman force on each atom is less than 0.0003 Hartree/Bohr, with the electronic energy convergence threshold set to 3.7 × 10$^{-6}$ Hartree. 
The bilayer distance between WSe$_{2}$ and graphene is found to be 3.47 {\AA}, which is in good agreement with previous studies \citep{Yang2018,Gmitra2016,Gmitra2017,Oh2024}. 
The interlayer distance reflects a weak van der Waals interaction, indicating that the interlayer coupling is minimal and the intrinsic electronic properties of graphene like linear dispersion near the Fermi level are well preserved in the heterostructure. 

\subsection{TDDFT}

We employ a TDDFT formalism for electron dynamics in presence of an electric field\cite{Bertsch2000,Otobe2008}.
We consider electron motion in a 2D material under irradiation of the electric field ${\bf E}(t)=-(1/c)d{\bf A}(t)/dt$  in the dipole approximation.
The orbital wavefunctions are defined in a box containing the unit cell of 2D material sandwiched between vacuum regions. 
The 2D material is assumed to be parallel to the $xy$ plane.
Each Bloch orbital $u_{n{\bf k}}({\bf r},t)$ is a two-component spinor, where $n$ and ${\bf k}$ denote the band index and 2D crystal momentum, respectively.
The time-dependent Kohn-Sham (TDKS) equation is described as follows:
\begin{equation}
\begin{split}i\hbar\frac{\partial}{\partial t}u_{n{\bf k}}({\bf r},t)=\Big[\frac{1}{2m}{\left(-i\hbar\nabla+\hbar{\bf k}+\frac{e}{c}{\bf A}(t)\right)}^{2}\\
-e\varphi({\bf r},t)+\hat{v}_{{\rm NL}}^{{{\bf k}+\frac{e}{\hbar c}{\bf A}(t)}}+{v}_{{\rm xc}}({\bf r},t)\Big]u_{n{\bf k}}({\bf r},t),
\end{split}
\label{eq:tdks}
\end{equation}
where the scalar potential $\varphi({\bf r},t)$ includes the Hartree
potential from the electrons and the local component of the ionic pseudopotentials and we have defined  $\hat{v}_{{\rm NL}}^{{\bf k}}\equiv e^{-i{\bf k}\cdot{\bf r}}\hat{v}_{{\rm NL}}e^{i{\bf k}\cdot{\bf r}}$. 
Here, $\hat{v}_{{\rm NL}}$ and ${v}_{{\rm xc}}({\bf r},t)$ represent the
nonlocal component of the ionic pseudopotentials and exchange-correlation
potential, respectively. 
We treat the dynamics of the valence electrons with the norm-conserving
pseudopotential \cite{Troullier1991} and assume the adiabatic LSDA for the exchange-correlation energy functional\cite{Perdew1981}.
SOC is incorporated through the $j$-dependent nonlocal potential $\hat{v}_{{\rm NL}}$ \cite{Theurich2001}.
Although the adiabatic LSDA exchange-correlation term is unable to describe spin torque effects, the SOC term in the nonlocal potential converts the angular momentum of excited carriers into spin-flipping torque~\cite{Neufeld2023}.

The $\alpha$-component ($\alpha=x,y,z$) of the spin magnetization per unit area is defined as
\begin{equation}
    m_{\alpha}(t) = \frac{1}{N_k} \sum_{{\bf k},n} f_{n{\bf k}} \int dz \int_{\Omega }\frac{dx dy}{\Omega } u^{\dagger}_{n{\bf k}}({\bf r},t) \frac{\sigma_{\alpha}}{2} u_{n{\bf k}}({\bf r},t),
    \label{eq:mag}
\end{equation}
where $\sigma_{\alpha}$ is the Pauli matrix and $\Omega$ is the area of the 2D unit cell.
$N_k$ is the number of k-points and $f_{n{\bf k}}$ is the occupation rate.
Based on the first order perturbative expansion of the wavefuction $u_{n{\bf k}}({\bf r},t)\approx u_{n{\bf k},t}({\bf r})+\delta u_{n{\bf k}}({\bf r},t)$,
where $u_{n{\bf k},t}({\bf r})$ is the adiabatic eigenstate and $\delta u_{n{\bf k}}({\bf r},t)$ represents the first-order correction, the first order contribution to 
 Eq.~(\ref{eq:mag}) is given by
\begin{equation}
 {\rm Re} [u^{\dagger}_{n{\bf k},t}({\bf r}) \sigma_{\alpha} \delta u_{n{\bf k}}({\bf r},t) ]\propto {\bf A},  
 \label{eq:fo_mag}
\end{equation}
while the second-order contribution is 
\begin{equation}
 \delta u^{\dagger}_{n{\bf k}}({\bf r},t) \frac{\sigma_{\alpha}}{2} \delta u_{n{\bf k}}({\bf r},t)\propto {\bf A}^2 .    
 \label{eq:so_mag}
\end{equation}

The ${\alpha\beta}$-component of the spin current density averaged over the unit cell is defined as
\begin{equation}
\begin{split}{J}^{\rm spin}_{\alpha\beta}(t)=\frac{1}{m}\int_{\Omega}\frac{d^3r}{\Omega}\frac{1}{N_k} {\rm Re} \sum_{{\bf k},n}f_{n{\bf k}} \, u_{n{\bf k}}^{\dagger}({\bf r},t) \sigma_{\alpha}\\
\times \left[
-i\hbar\nabla_{\beta}+\hbar{k}_{\beta}+\frac{e}{c}{A}_{\beta}(t)
\right]
u_{n{\bf k}}({\bf r},t) +{J}^{\rm spin}_{\rm NL,\alpha\beta}(t),
\end{split}
\end{equation}
where ${J}^{\rm spin}_{\rm NL,\alpha\beta}(t)$ is a correction term that stems from the non-local potential:
\begin{equation}
\begin{split}
{J}^{\rm spin}_{\rm NL,\alpha\beta}(t)=\int_{\Omega}\frac{d^3r}{\Omega}\frac{1}{N_k} {\rm Re} \sum_{{\bf k},n}f_{n{\bf k}} \, u_{n{\bf k}}^{\dagger}({\bf r},t) \sigma_{\alpha}\\
\times \frac{1}{i\hbar}\left[{r}_{\beta},\hat{v}_{{\rm NL}}^{{{\bf k}+\frac{e}{\hbar c}{\bf A}(t)}}\right]
u_{n{\bf k}}({\bf r},t).
\end{split}
\end{equation}
Without the Pauli matrix $\sigma_{\alpha}$, the above definition is equivalent to the ${\beta}$-component of the electron current density averaged over the unit cell, ${J_{\beta}}(t)$.

The occupancy of the time-dependent wavefunctions at time $t$ is estimated by projecting $\{u_{n{\bf k}}(t)\}$ onto the ground-state orbitals $\{u^{\rm GS}_{n{\bf k}}\}$ as
\begin{equation}
    F_{n{\bf k}}(t) =  \sum_{m} f_{m{\bf k}} \left| \left\langle u^{\rm GS}_{n{\bf k}} \middle| u_{m{\bf k}}(t) \right\rangle \right|^2,
    \label{eq:occ}
\end{equation}
At $t=0$, $F_{n{\bf k}}(t)$ reduces to the ground-state occupancy $f_{n{\bf k}}$.
To visualize changes in occupancy, we define the weighted local density of states as
\begin{equation}
    w(\varepsilon,{\bf r},t) = \frac{1}{N_k} \sum_{{\bf k},n} \Delta F_{n{\bf k}}(t) \left| u^{\rm GS}_{n{\bf k}}({\bf r}) \right|^2 \delta(\varepsilon - \varepsilon^{\rm GS}_{n{\bf k}}),
    \label{eq:pdos}
\end{equation}
where $\Delta F_{n{\bf k}}(t) = F_{n{\bf k}}(t) - f_{n{\bf k}}$.
This quantity represents the density distribution of excited carriers at time $t$.
In the absence of $\Delta F_{n{\bf k}}(t)$, the above expression is equivalent to the conventional local density of states (LDoS), $D(\varepsilon,{\bf r})$.
By sorting each component with respect to $(n,{\bf k})$, the spin-up and spin-down contributions can be separated as $w_{\uparrow,\downarrow}(\varepsilon,{\bf r},t)$.

For the atomic structure of the WSe$_2$-graphene HB obtained in Sec.~\ref{sec:structure}, we will perform first-principles TDDFT calculations.
In this paper, TDDFT calculations are carried out using SALMON code\cite{SALMON_web,Noda2019}.
The norm-conserving pseudopotentials are the same as those used in the structure optimization via the OpenMX code.
The spatial grid sizes and k-points are optimized according to the converging results. 
For both the conventional WSe$_2$ monolayer and WSe$_2$-graphene HB, the determined parameter of the real-space grid spacing is 0.21 {\AA}.
The optimized k-mesh in the 2D Brillouin zone is $16\times 16$ for the conventional WSe$_2$ monolayer and $8\times 8$ for the WSe$_2$-graphene HB. 
The time step is set to $dt=5\times 10^{-4}$ fs.
The calculation conditions for the conventional WSe$_2$ monolayer are almost the same as those employed in Refs.~\onlinecite{Hashmi2022a,Hashmi2022,Hashmi2023,Yamada2023}.

The vector potential of the applied circularly polarized light pulse is given by
\begin{eqnarray}
{\bf A}(t) &=& -\frac{cE_{\rm max}}{\omega} \sin^{4}\left(\frac{\pi t}{T}\right) \nonumber \\
&\times&
\left[ \hat{{\bf x}} \cos\left\{ \omega\left(t - \frac{T}{2}\right) \right\} + \hat{{\bf y}} \sin\left\{ \omega\left(t - \frac{T}{2}\right) \right\} \right], \nonumber \\
&& (0 < t < T),
\end{eqnarray}
where $\hbar \omega = 1.5$~eV and $T = 20$~fs.  
The peak electric field amplitude $E_{\rm max}$ is chosen to yield a peak intensity ranging from $I = 10^{8}$ to $10^{12}$~W/cm$^2$.

\section{Results}

\begin{figure}
    \begin{tabular}{c}
    \includegraphics[keepaspectratio,width=\columnwidth]{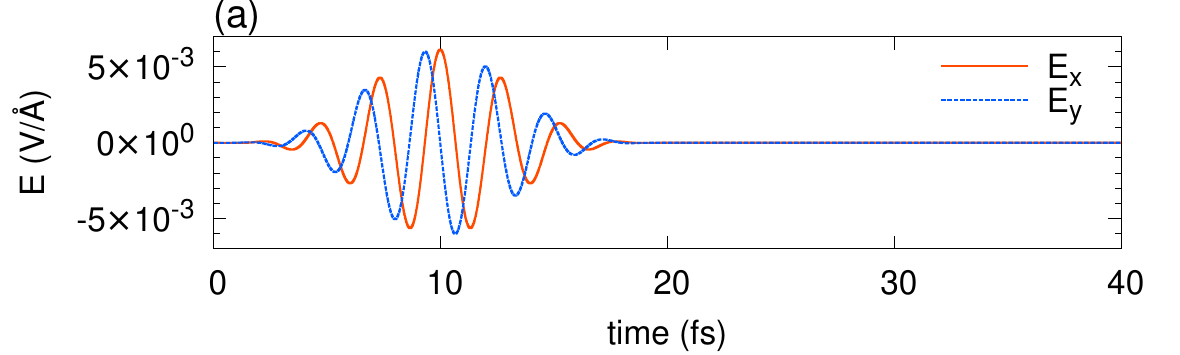}\\
    \includegraphics[keepaspectratio,width=\columnwidth]{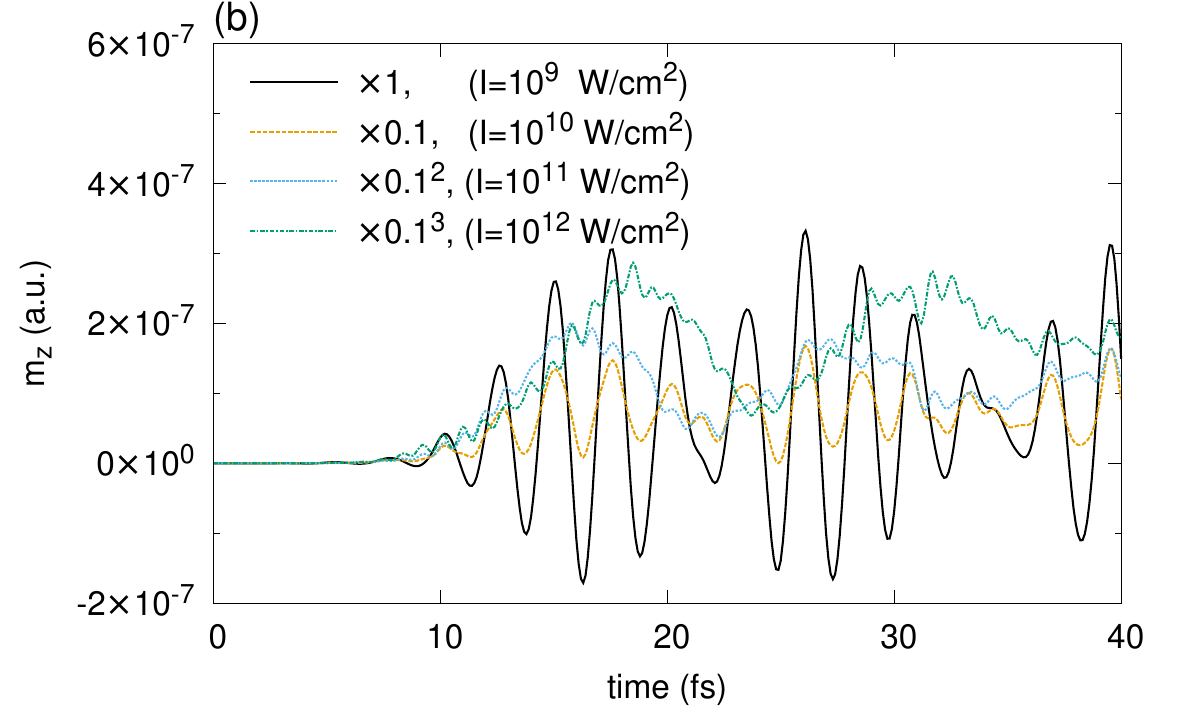}\\
    \includegraphics[keepaspectratio,width=\columnwidth]{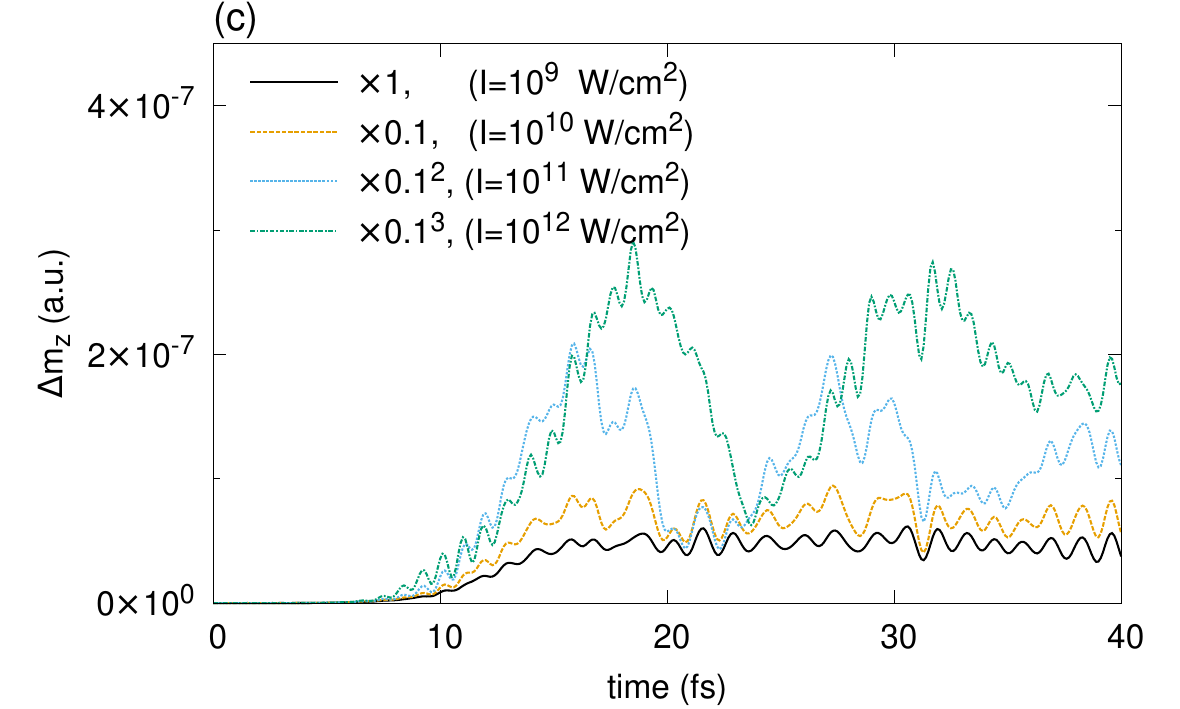}
    \end{tabular}
    \caption{\label{fig:wse2_spin}
    (a) Applied electric field.
    (b) Spin magnetization per unit area in the conventional WSe$_2$ monolayer compared across a range of laser intensities $I$ from $10^9$ to $10^{12}$ W/cm$^2$, scaled by a factor of $0.1^n$ ($n = 0, 1, 2, 3$ for $I = 10^{9+n}$~W/cm$^2$). 
    (c) Scaled nonlinear contribution of (b) obtained by subtracting the scaled linear response.
    }
\end{figure}

Before discussing heterostructures, we consider ultrafast spin dynamics in the conventional WSe$_2$ monolayer under circularly polarized laser pulse.
Figure~\ref{fig:wse2_spin} shows the calculated spin magnetization per unit area in monolayer WSe$_2$ under pulsed electric field excitation. 
Panel (a) is the applied electric field with an intensity of $I=10^9$ W/cm$^2$, while panel (b) shows the resulting spin magnetization $m_z(t)$. 
Panel (c) isolates the contribution from nonlinear components by subtracting the scaled linear response. 
In panel (b), the spin magnetization is compared across a range of laser intensities from $10^9$ to $10^{12}$ W/cm$^2$.
Because the real excitation of the spin magnetization is expected to be proportional to the excited carrier density, the lines in panel (b) are scaled by a factor of $0.1^n$ ($n = 0, 1, 2, 3$ for $I = 10^{9+n}$~W/cm$^2$) for comparison.
In the low-intensity regime, the nodal structure for $10^9$ W/cm$^2$ (black solid line) coincides with that for $10^{10}$ W/cm$^2$ (orange dashed line), but the former exhibits a pronounced oscillation with the same frequency as that of the applied electric field.
This indicates that the spin magnetization in this regime is a superposition of components proportional to the first- and second-order terms of the electric field. 
The second-order component corresponds to the real excitation of the spin magnetization.

As the intensity increases, the contribution from the linear (first-order) term diminishes its importance.
To isolate the nonlinear response, panel (c) subtracts the scaled linear component obtained from the $10^{8}$ W/cm$^2$ result and multiplied by $\sqrt{10}^{n+1}$ from each spin magnetization, $\Delta m_z(t) = m_z(t) - \sqrt{10}^{n+1} m_z(t)|_{I=10^8\,{\rm W/cm^2}}$. 
This procedure reveals the behavior of the higher-order terms of the spin magnetization.
As expected, the curves for $10^{9}$ and $10^{10}$ W/cm$^2$ closely overlap. 
At higher intensities, nonlinear excitation involving higher-order terms becomes significant. 
The behavior suggests that the spin magnetization scales approximately with the square of the electric field, in accordance with Eq~(\ref{eq:so_mag}).

\begin{figure}
    \begin{tabular}{c}
    \includegraphics[keepaspectratio,width=\columnwidth]{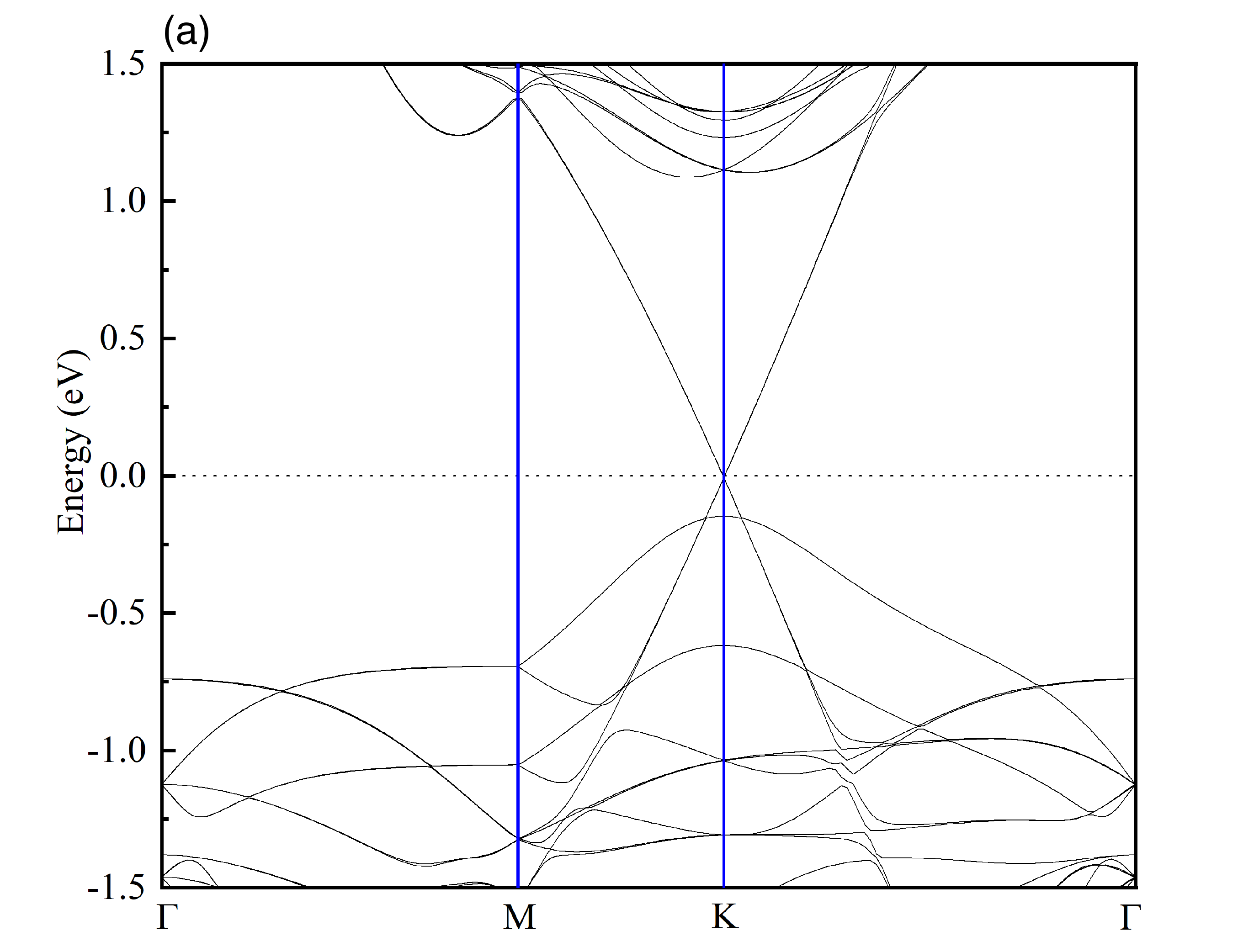}\\
    \includegraphics[keepaspectratio,width=\columnwidth]{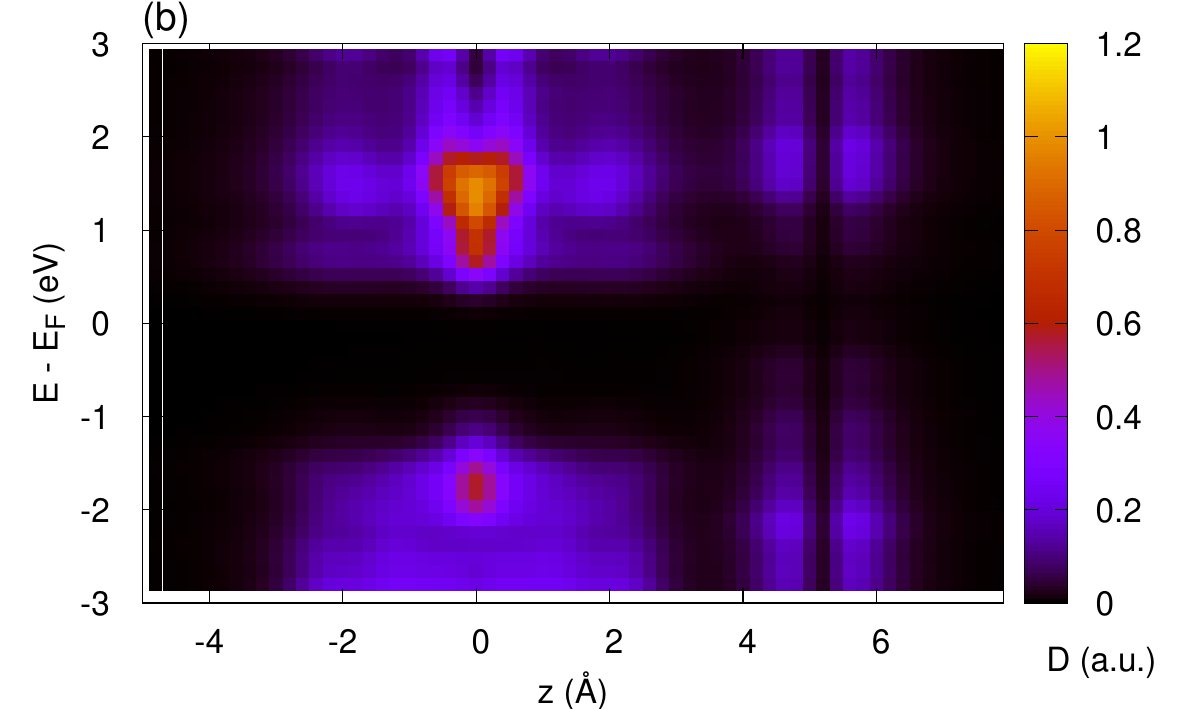}\\
    \includegraphics[keepaspectratio,width=\columnwidth]{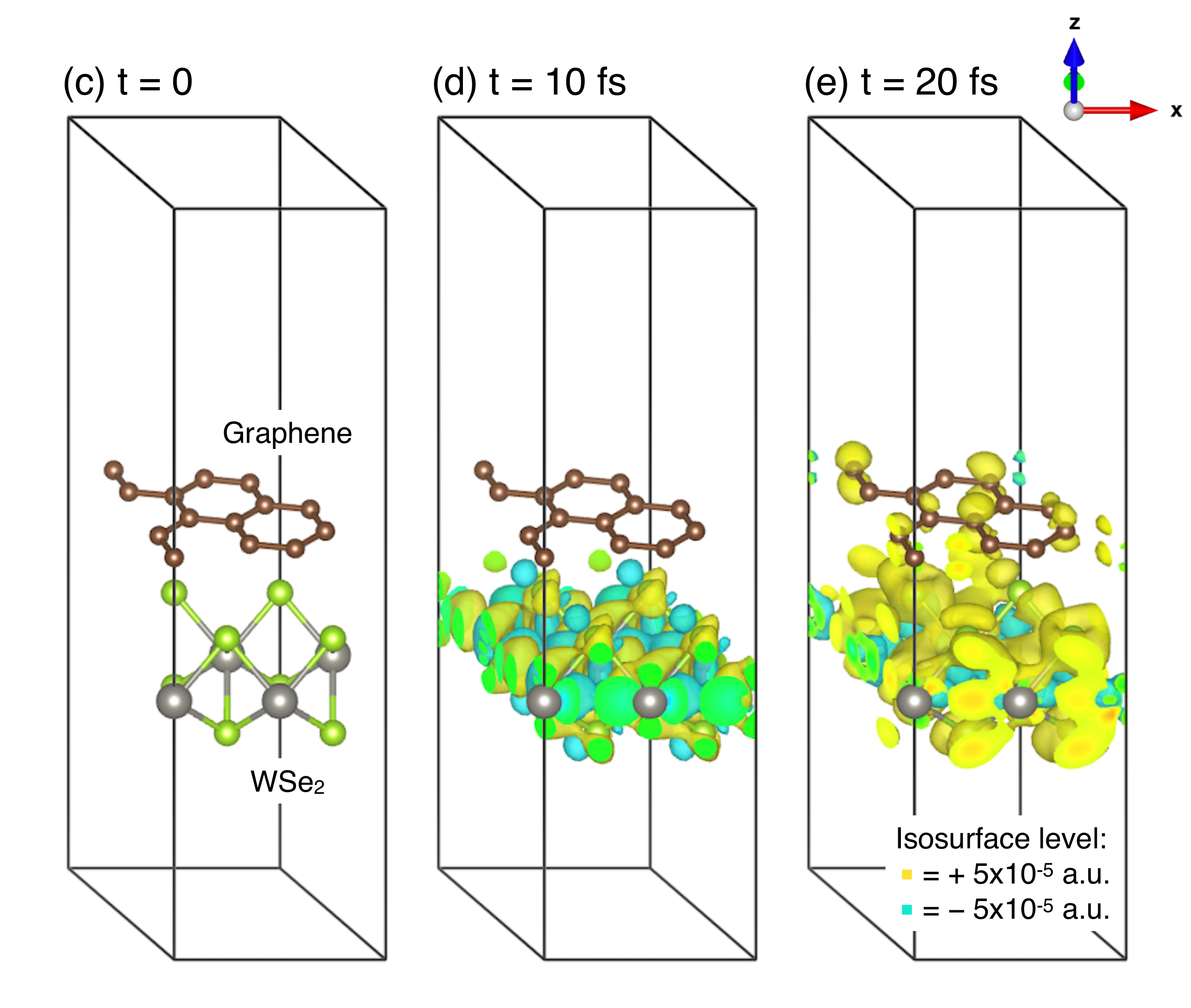}
    \end{tabular}
    \caption{\label{fig:band}
    (a) Band structure of the WSe$_2$–graphene HB.
    (b) Local density of states.
    (c--e) Snapshots of the spin magnetization density distribution from $t=0$ to $t=$ 20 fs under laser irradiation with $I = 10^{12}$ W/cm$^2$ (visualized by VESTA\cite{Momma2011}).
    }
\end{figure}

Next, we investigate the WSe$_2$–graphene HB.  
Figures~\ref{fig:band}(a) and (b) show the band structure and LDoS, respectively, for the atomic structure obtained through structural optimization [Fig.~\ref{fig:band}(c)].  
In panel (b), the origin of the $z$-axis (horizontal axis) is set at the position of the tungsten (W) layer.  
The selenium (Se) layers are located at $z = \pm 1.7$~\AA.  
At $z = 0$, the conduction band minimum and valence band maximum of WSe$_2$ appear around 0.5 eV and $-1$ eV, respectively.  
The graphene layer is positioned at $z = 5.2$~\AA, and its Dirac cone is centered around the Fermi energy ($=0$).
Panels (c), (d) and (e) present snapshots of the density distribution of the spin magnetization [integrand of Eq.~(\ref{eq:mag})] at time $t=0$, 10, and 20 fs, respectively, under laser irradiation with $I = 10^{12}$ W/cm$^2$.

\begin{figure}
    \begin{tabular}{c}
    \includegraphics[keepaspectratio,width=\columnwidth]{fig_pulse.pdf}\\    
    \includegraphics[keepaspectratio,width=\columnwidth]{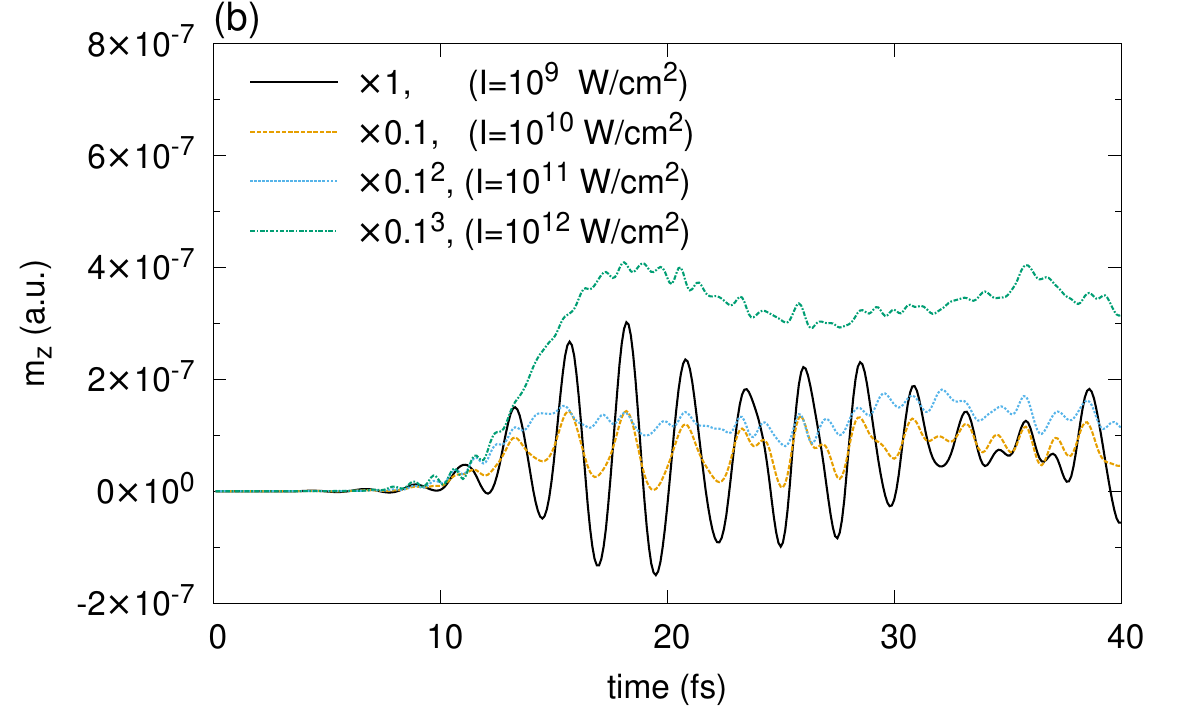}\\
    \includegraphics[keepaspectratio,width=\columnwidth]{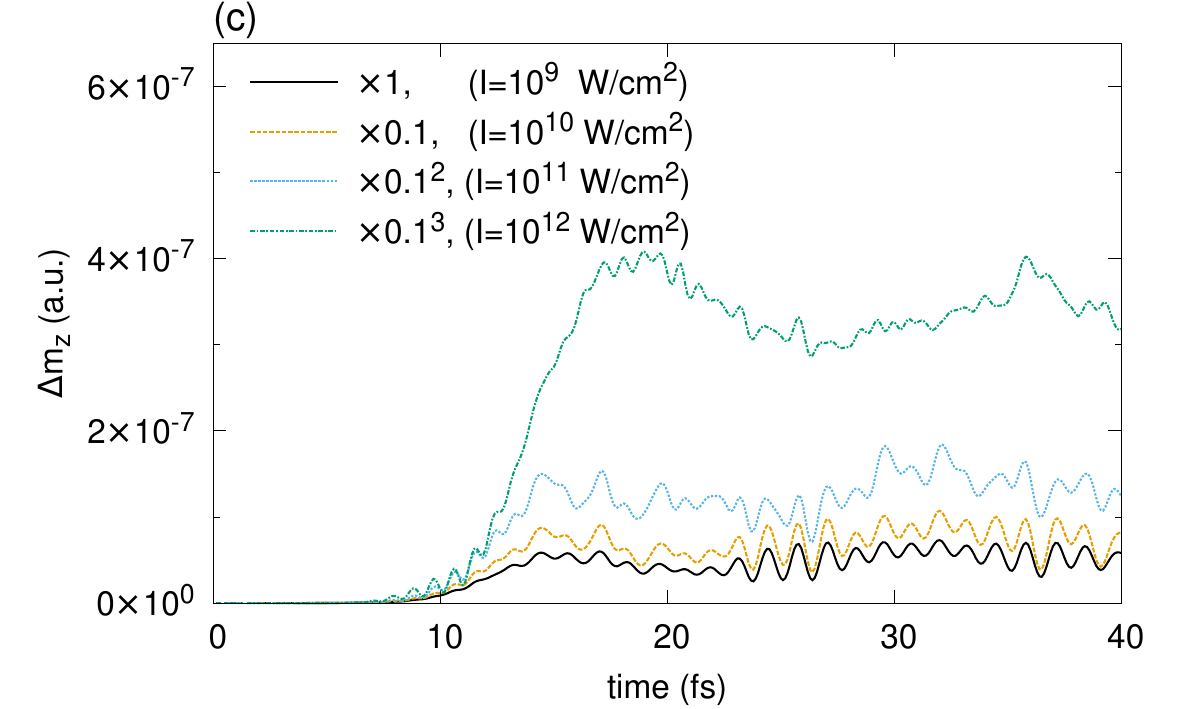}
    \end{tabular}
    \caption{\label{fig:hetero_spin}
    The same as Fig.~\ref{fig:wse2_spin} but for the WSe$_2$–graphene HB.
    }
\end{figure}

Figure~\ref{fig:hetero_spin} presents the same analysis as Fig.~\ref{fig:wse2_spin}, but for the WSe$_2$–graphene HB.  
Panels (b) and (c) show the total spin magnetization per unit area of the HB.  
The behavior of the total spin magnetization is qualitatively similar to that of the WSe$_2$ monolayer, but the results at higher intensities exhibit a slight increase due to changes in the density of states (DoS) of the WSe$_2$ layer induced by the heterojunction.
This DoS change originates mainly from the cell compression of the WSe$_2$ layer in the $xy$ plane.

\begin{figure}
    \begin{tabular}{c}
    \includegraphics[keepaspectratio,width=\columnwidth]{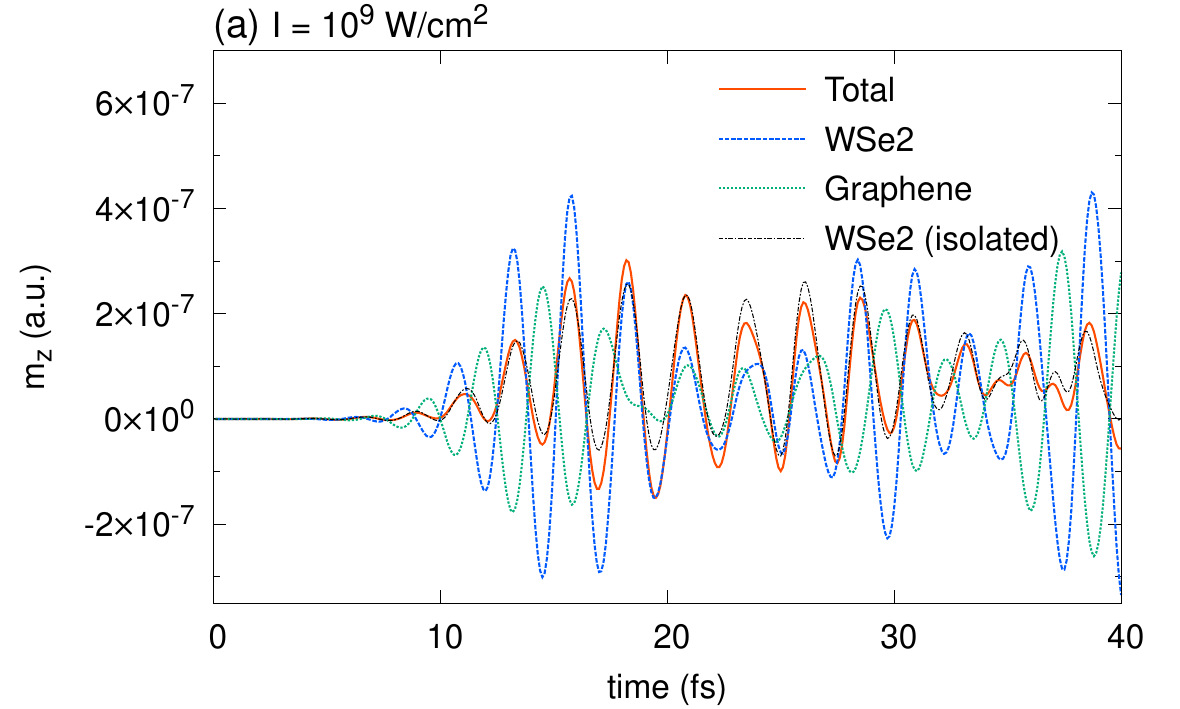}\\
    \includegraphics[keepaspectratio,width=\columnwidth]{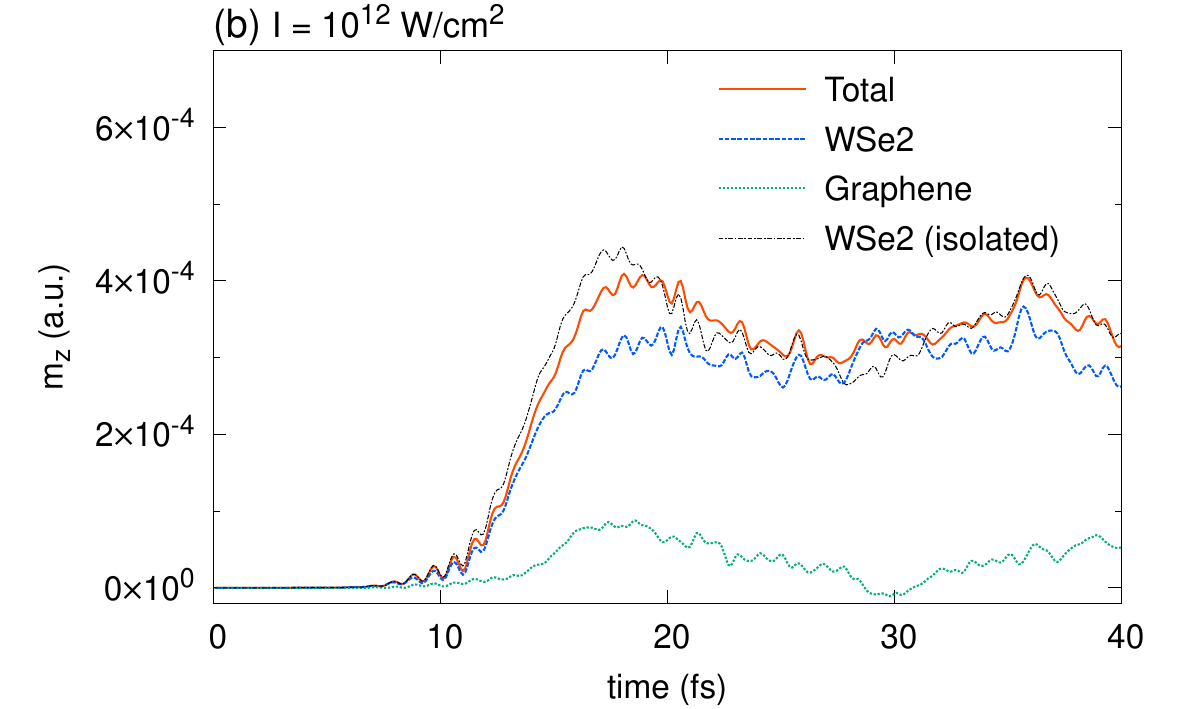}
    \end{tabular}
    \caption{\label{fig:hetero_spin_decomp}
    Total spin magnetization $m_{z}(t)$ and its decomposition into contributions from each layer under (a) $I=10^{9}$ W/cm$^2$ and (b) $I=10^{12}$ W/cm$^2$.
    }
\end{figure}

\begin{figure}
    \begin{tabular}{c}
    \includegraphics[keepaspectratio,width=\columnwidth]{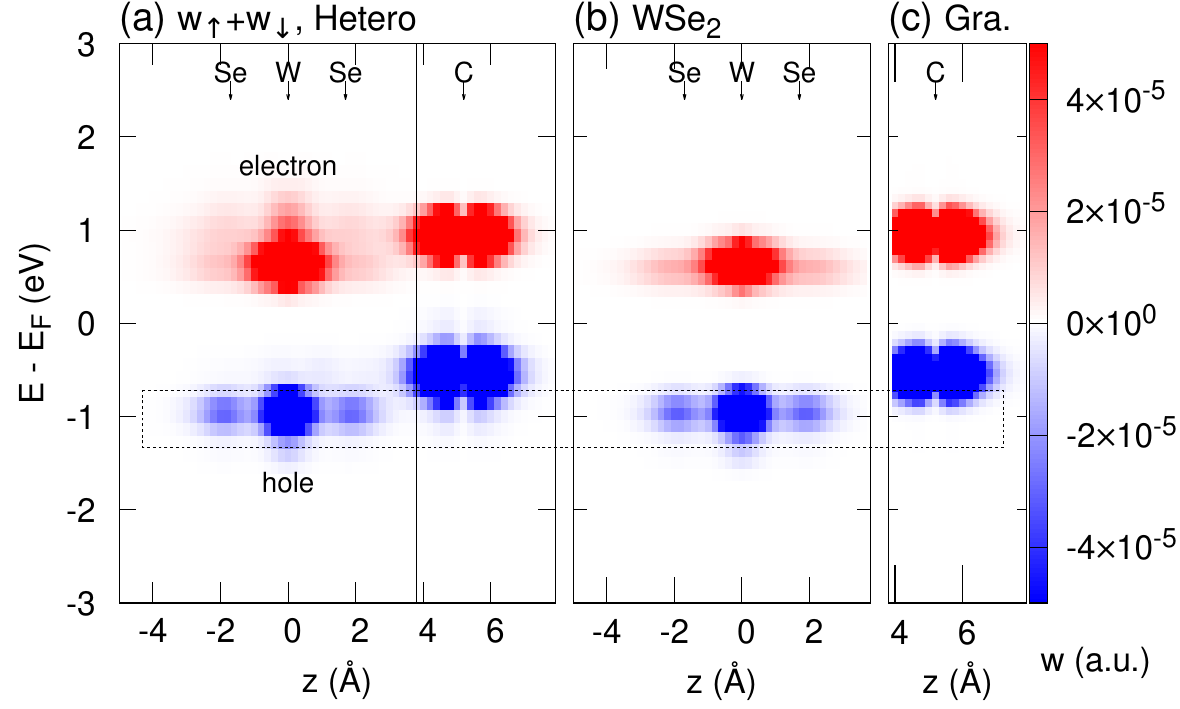}\\
    \includegraphics[keepaspectratio,width=\columnwidth]{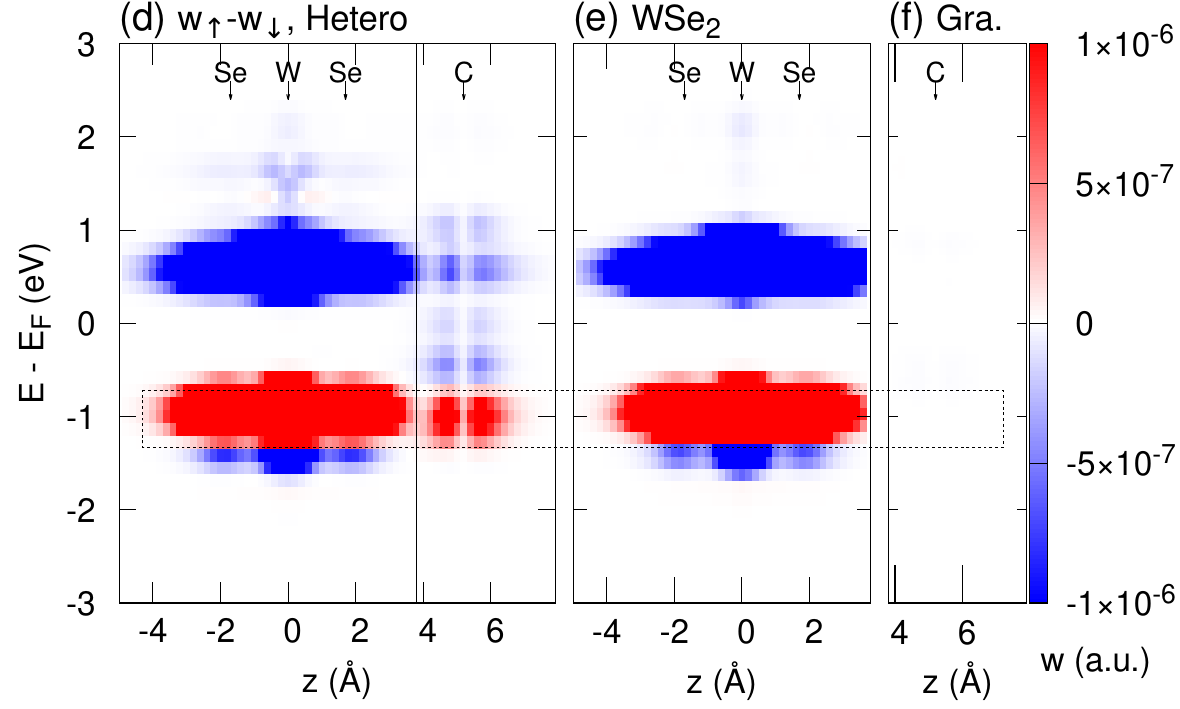}
    \end{tabular}
    \caption{\label{fig:ldos_int1e9}
    Local density of states weighted by the change in occupancy from the ground state, $w_{\uparrow,\downarrow}(\varepsilon,z)$, at the pulse end with an intensity of $I = 10^{9}$ W/cm$^2$.
    (a--c) [(d--f)] Sum (Difference) of the spin-up and spin-down contributions.
    Panels (a) and (d) correspond to the HB results, while (b) and (e) [(c) and (f)] show the results for the isolated WSe$_2$ (graphene) monolayer.  
    }
\end{figure}

In Fig.~\ref{fig:hetero_spin_decomp}(a) [(b)], we compare the total spin magnetization and its decomposition into contributions from each layer under $I=10^{9}$ W/cm$^2$ ($I=10^{12}$ W/cm$^2$).  
The red solid lines correspond to those shown in Fig.~\ref{fig:hetero_spin}(b).  
The blue dashed and green dotted lines represent the spin magnetization integrated over the WSe$_2$ and graphene regions, respectively, as separated by the vertical line in Fig.~\ref{fig:ldos_int1e9}(a).  
The black dash-dotted lines indicate the results for the isolated WSe$_2$ monolayer, calculated using the same atomic positions and supercell as in the HB case.
From Fig.~\ref{fig:hetero_spin_decomp}(a, b), the total spin magnetization of the HB almost coincides with that of the isolated WSe$_2$ monolayer.  
A portion of the spin magnetization induced in the WSe$_2$ layer is transferred to the graphene layer.
As mentioned above, the spin magnetization of the isolated WSe$_2$ monolayer (black dash-dotted line) is slightly modified from that of the conventional WSe$_2$ monolayer (Fig.~\ref{fig:wse2_spin}) due to the DoS change caused by cell compression.

\begin{figure*}
    \begin{tabular}{cc}
    
    \begin{minipage}[t]{\columnwidth}
    \includegraphics[keepaspectratio,width=\textwidth]{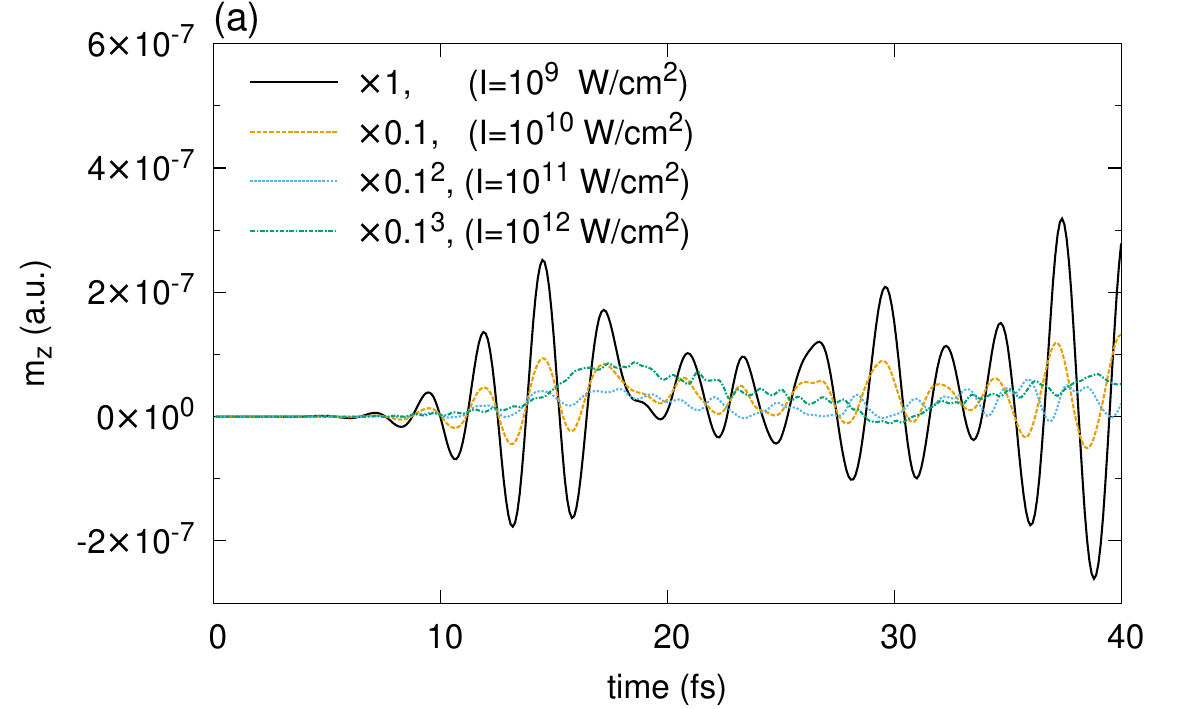}
    \end{minipage} &
    
    \begin{minipage}[t]{\columnwidth}
    \includegraphics[keepaspectratio,width=\textwidth]{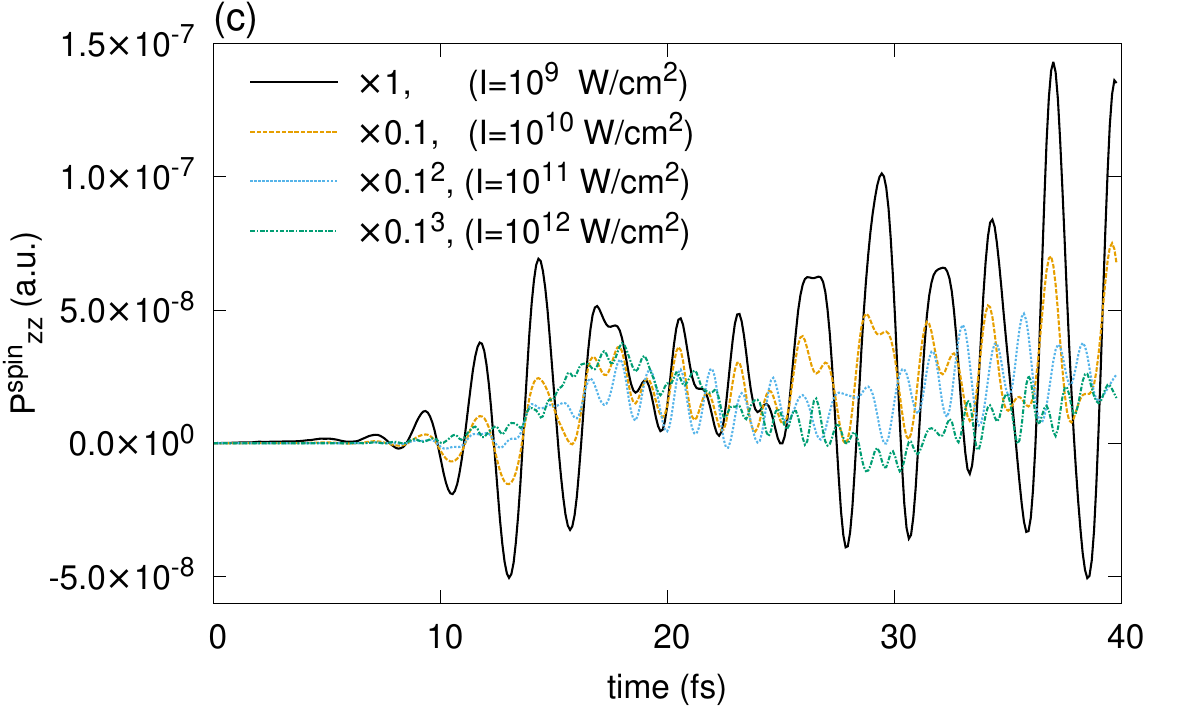}
    \end{minipage} \\

    \begin{minipage}[t]{\columnwidth}
    \includegraphics[keepaspectratio,width=\textwidth]{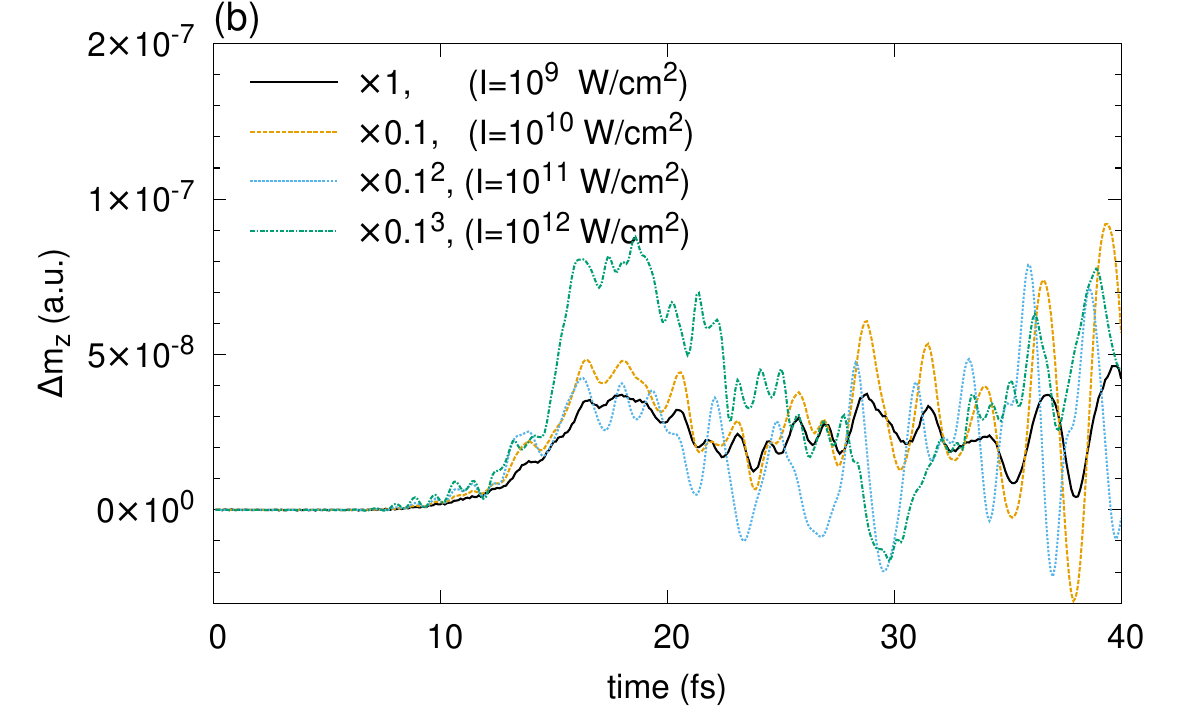}
    \end{minipage} &
    
    \begin{minipage}[t]{\columnwidth}
    \includegraphics[keepaspectratio,width=\textwidth]{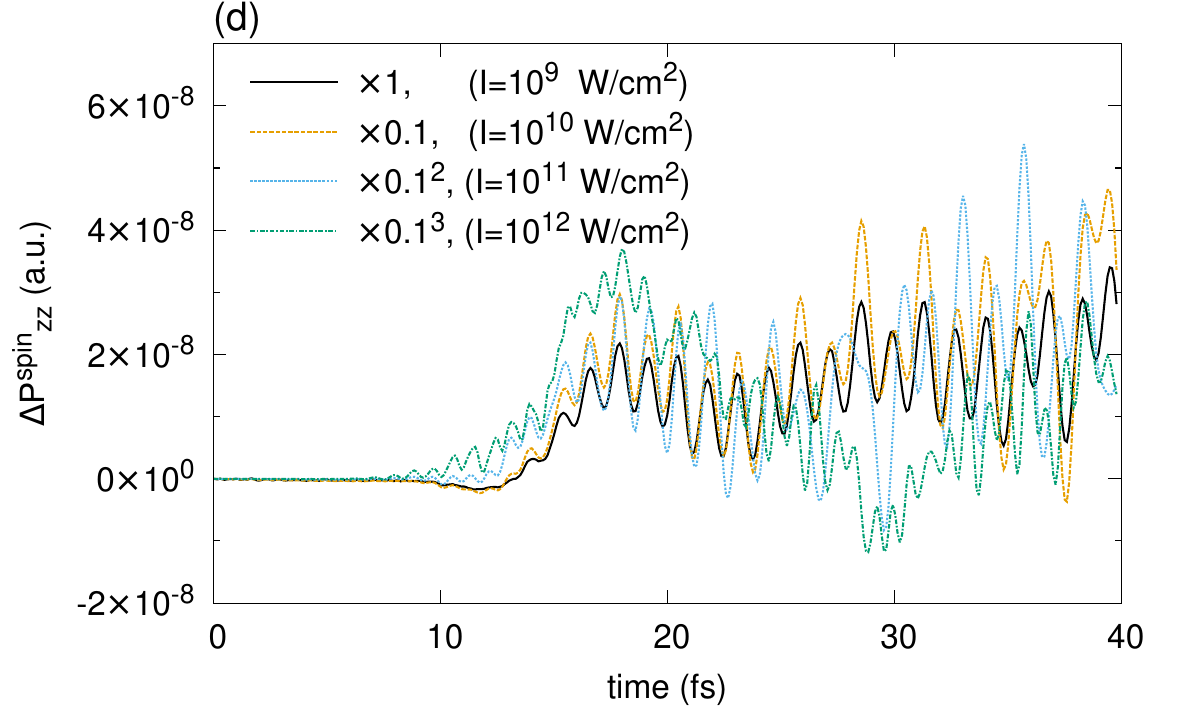}
    \end{minipage}
    
    \end{tabular}
    \caption{\label{fig:hetero_spin_polarization}
    Various quantities in the WSe$_2$–graphene HB under laser intensities $I$ ranging from $10^9$ to $10^{12}$~W/cm$^2$.  
    All curves are scaled by a factor of $0.1^n$ ($n = 0, 1, 2, 3$ for $I = 10^{9+n}$~W/cm$^2$) for comparison.
    (a) Spin magnetization around graphene.
    (b) Nonlinear component of spin magnetization around graphene.
    (c) Spin transfer $P^{\rm spin}_{zz}(t)$.
    (d) Nonlinear component of spin transfer ${\Delta}P^{\rm spin}_{zz}(t)$.
    }
\end{figure*}

The intensity dependence of the transferred spin magnetization around the graphene layer (green dotted lines in Fig.~\ref{fig:hetero_spin_decomp}) is shown in Fig.~\ref{fig:hetero_spin_polarization}(a, b).  
Panels (a) and (b) display the scaled spin magnetization of graphene and its nonlinear component, respectively, following the same procedure as in Figs.~\ref{fig:hetero_spin}(b) and (c).  
Figures~\ref{fig:hetero_spin_polarization}(c) and (d) are analogous to panels (a) and (b), but represent the time-integrated $zz$-component of the averaged spin current density, $P^{\rm spin}_{zz}(t)=\int_0^t dt'\, J^{\rm spin}_{zz}(t')$, where panel (d) shows $\Delta P^{\rm spin}_{zz}(t) = P^{\rm spin}_{zz}(t) - \sqrt{10}^{n+1} P^{\rm spin}_{zz}(t)\big|_{I=10^8\,{\rm W/cm^2}}$.
This quantity represents the spin transfer, or the polarization of the $z$-component spin along the $z$-axis.  

By comparing panels (a) and (c), as well as (b) and (d), we find that the behavior of the graphene spin magnetization and the spin transfer along the $z$-axis are qualitatively similar.  
These results confirm that the spin transfer from the WSe$_2$ layer to the graphene layer is roughly proportional to the square of the electric field, consistent with the spin magnetization observed in the isolated WSe$_2$ monolayer.  
The spin transfer occurs through the migration of a portion of the spin magnetization from the WSe$_2$ layer to the graphene layer.

\begin{figure}
    \includegraphics[keepaspectratio,width=\columnwidth]{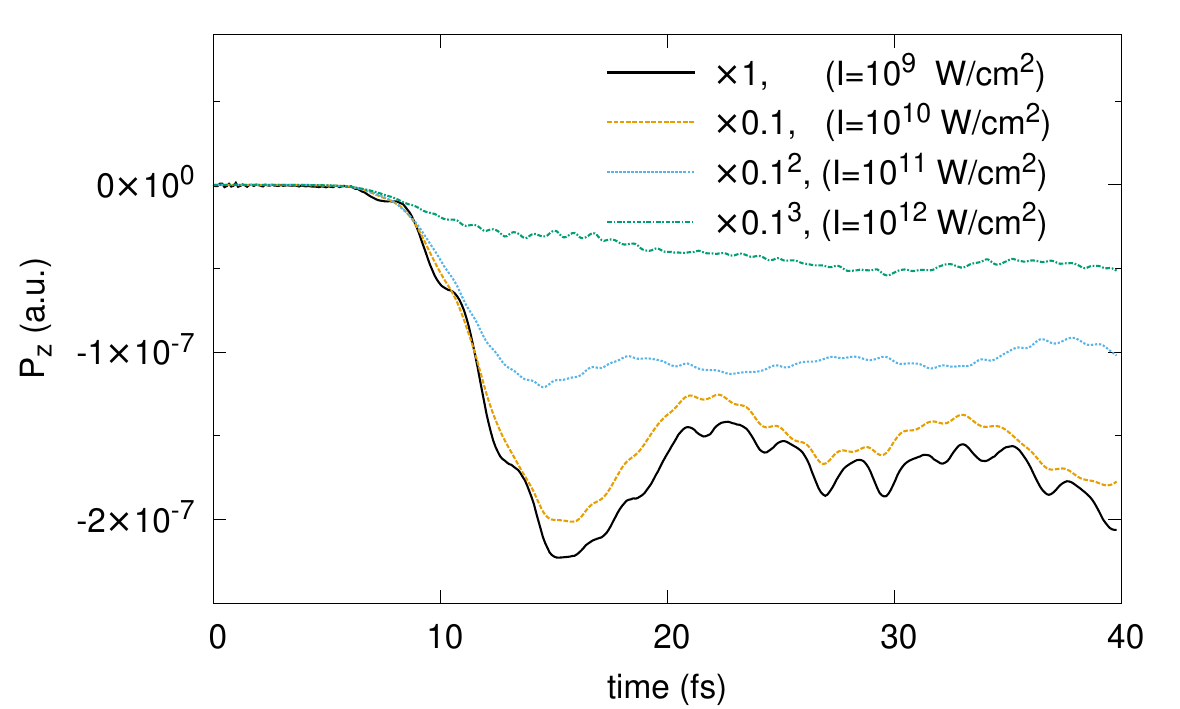}
    \caption{\label{fig:polarization} 
    Scaled electron transfer in the WSe$_2$–graphene HB along the $z$-axis, $P_{z}(t)$, compared across a range of laser intensities from $10^9$ to $10^{12}$ W/cm$^2$. 
    }
\end{figure}

Next, we consider charge transfer in the HB.
Figure~\ref{fig:polarization} shows the intensity dependence of the polarization along the $z$-axis, defined as $P_z(t) = \int_0^t dt'\, J_z(t')$, where ${\bf J}(t)$ denotes the electron current density.  
This quantity represents the amount of electron transfer between the two layers, analogous to the spin transfer shown in Fig.~\ref{fig:hetero_spin_polarization}(c).  
The negative sign of the curves in this figure indicates that electrons are transferred from the graphene layer to the WSe$_2$ layer.
This observation is consistent with the previous first-principles study for a MoS$_2$–graphene HB under a linearly-polarized laser pulse~\cite{Iida2018}. 
Electrons are transferred from the graphene layer to the TMD layer due to the DoS imbalance between the two layers.
In the low-intensity regime, the polarization is proportional to the square of the electric field, as discussed in Ref.~\onlinecite{Iida2018}.  
At higher intensities, the polarization saturates with increasing the intensity, which can be attributed to the Pauli blocking by excited and transferred carriers.
The positive sign of the curves in Fig.~\ref{fig:hetero_spin_polarization}(c) indicates that spin transfer occurs in the opposite direction to the electron transfer, namely from WSe$_2$ to graphene.

Figure~\ref{fig:ldos_int1e9} shows the weighted LDoS $w_{\uparrow,\downarrow}(\varepsilon,z)$ [Eq.~(\ref{eq:pdos})], as a function of the orbital energy $\varepsilon$ and $z$ coordinate, at the pulse end ($t = 20$~fs) with $I = 10^9$~W/cm$^2$.  
Panels (a--c) [(d--f)] show the sum (difference) of the spin-up and spin-down contributions, $w_{\uparrow}+w_{\downarrow}$ ($w_{\uparrow}-w_{\downarrow}$).  
In panels (a--c) [(d--f)], the positive and negative regions represent the spatial distributions of excited electrons and holes (up-spin and down-spin polarizations), respectively. 
Panels (a) and (d) correspond to the HB results, while (b) and (e) [(c) and (f)] show the results for the isolated WSe$_2$ (graphene) monolayer. 
At this low laser intensity, the excitation is expected to remain within the linear response regime. 

In Figs.~\ref{fig:ldos_int1e9}(a--c), the result for the HB (a) shows strong similarity to the superposition of the isolated monolayer results (b, c), but the hot spots of excited carriers differ along the energy axis.
Around $z = 0$ and $\varepsilon=1$ eV in Fig.~\ref{fig:ldos_int1e9}(a), excited electrons are observed that are absent in Fig.~\ref{fig:ldos_int1e9}(b).  
These electrons are transferred from the graphene layer to the WSe$_2$ layer due to the band offset and the DoS imbalance between the two layers~\cite{Iida2018}.  
Regarding the spin imbalance of excited carriers, Fig.~\ref{fig:ldos_int1e9}(d) clearly shows spin transfer from the WSe$_2$ layer to the graphene layer.
As expected, the isolated graphene does not exhibit spin excitation [Fig.~\ref{fig:ldos_int1e9}(f)].
From Fig.~\ref{fig:ldos_int1e9}(d), the spin distribution in the graphene layer corresponds to that in the WSe$_2$ layer.

\begin{figure}
    \begin{tabular}{c}
    \includegraphics[keepaspectratio,width=\columnwidth]{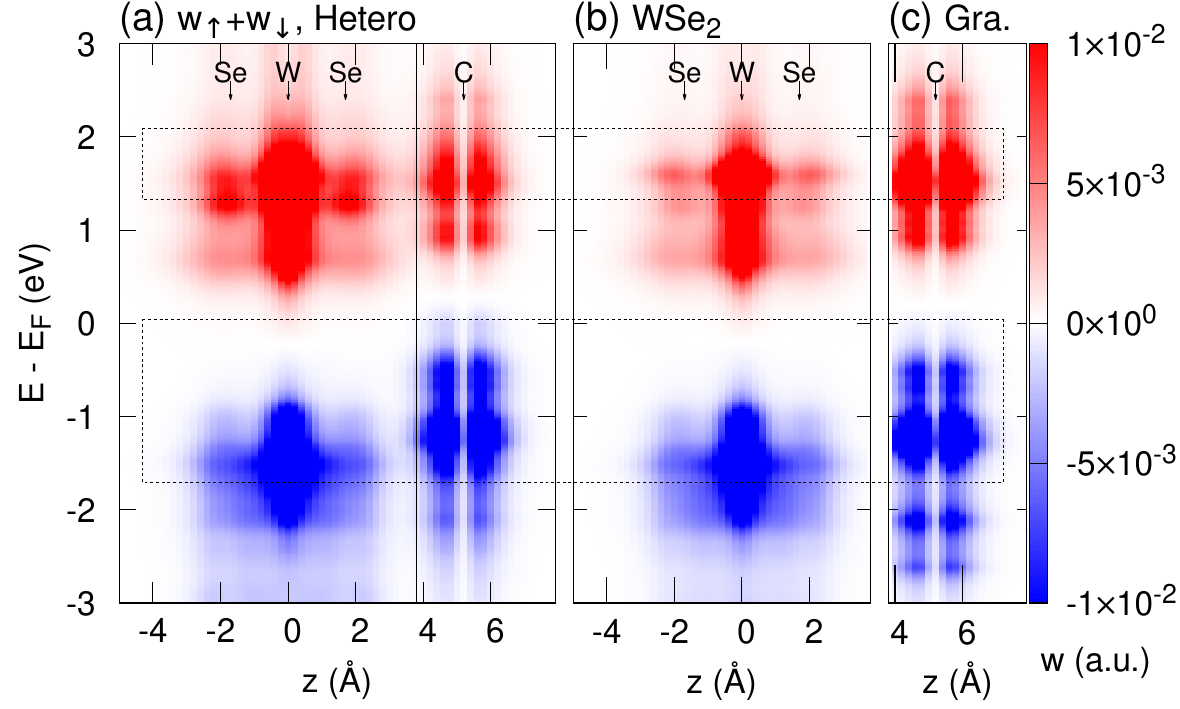}\\
    \includegraphics[keepaspectratio,width=\columnwidth]{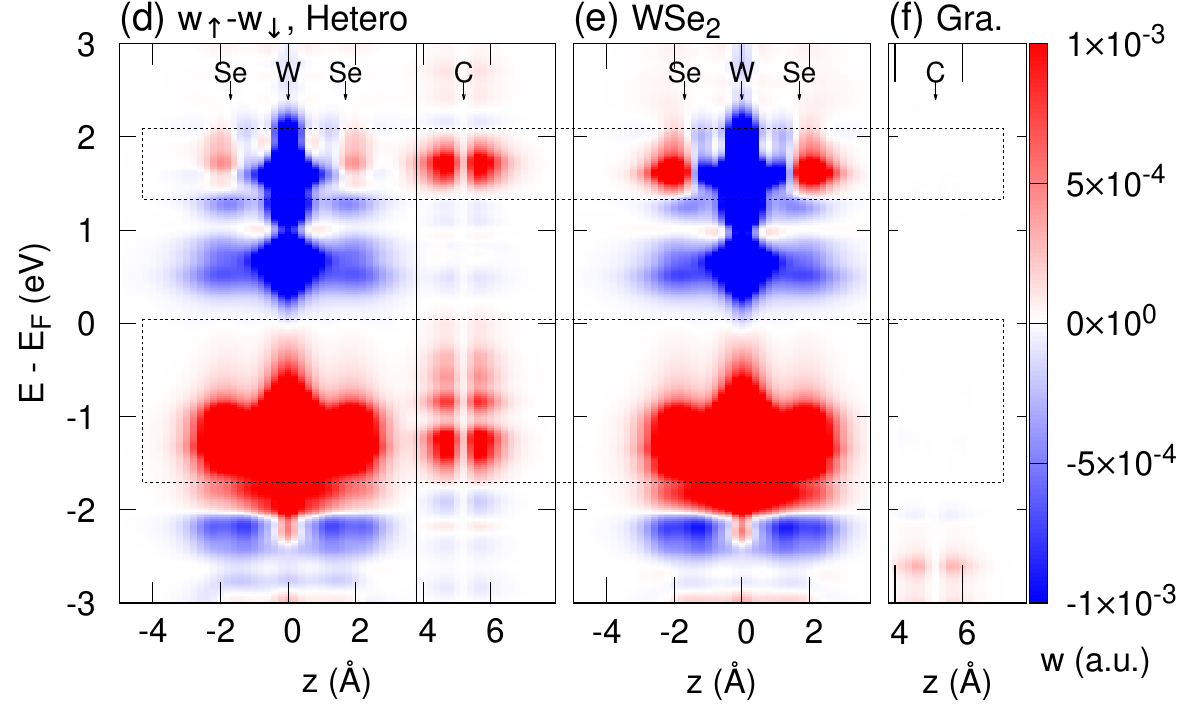}
    \end{tabular}
    \caption{\label{fig:ldos_int1e12}
    The same as Fig.~\ref{fig:ldos_int1e9} but with $I = 10^{12}$ W/cm$^2$.
    }
\end{figure}

Figure~\ref{fig:ldos_int1e12} presents the same analysis as Fig.~\ref{fig:ldos_int1e9}, but under a higher laser intensity of $I = 10^{12}$ W/cm$^2$.  
At this intensity, excited carriers are distributed over a broader energy range due to nonlinear multiphoton excitation.  
While the overall trend remains similar to that in Fig.~\ref{fig:ldos_int1e9}, the spin distribution shown in Fig.~\ref{fig:ldos_int1e12}(d, e) becomes more complex.  
The spin distribution around the Se layer at $z = 1.7$~{\AA} appears to be transferred to the graphene layer, although it differs from the distribution observed near the W layer.
In both of Fig.~\ref{fig:ldos_int1e9}(d) and Fig.~\ref{fig:ldos_int1e12}(d), the spin distribution in the graphene layer corresponds to that in the Se layer at $z = 1.7$~{\AA}.

We focus on the migration of up-spin electrons because it is the main contribution to the spin transfer.
The boxed areas outlined by the black dotted lines in Figs.~\ref{fig:ldos_int1e9} and \ref{fig:ldos_int1e12} highlight the regions where the spin-up polarization emerges in the graphene layer of the HB.
The horizontal ($z$-axis) transport of carriers within these regions gives rise to a net positive spin current from WSe$_2$ to graphene, as indicated by the positive spin transfer in Fig.~\ref{fig:hetero_spin_polarization}. 
A comparison of the vicinity of graphene in Figs.~\ref{fig:ldos_int1e9}(a) and (c), as well as in Figs.~\ref{fig:ldos_int1e12}(a) and (c), within the boxed areas, reveals the corresponding carrier densities exhibit a decrease in electron density and an increase in hole density in the graphene layer of the HB. 
This observation suggests that electrons are transferred from graphene to WSe$_2$, which is consistent with the negative electron transfer shown in Fig.~\ref{fig:polarization}.

\begin{figure}
    \includegraphics[keepaspectratio,width=\columnwidth]{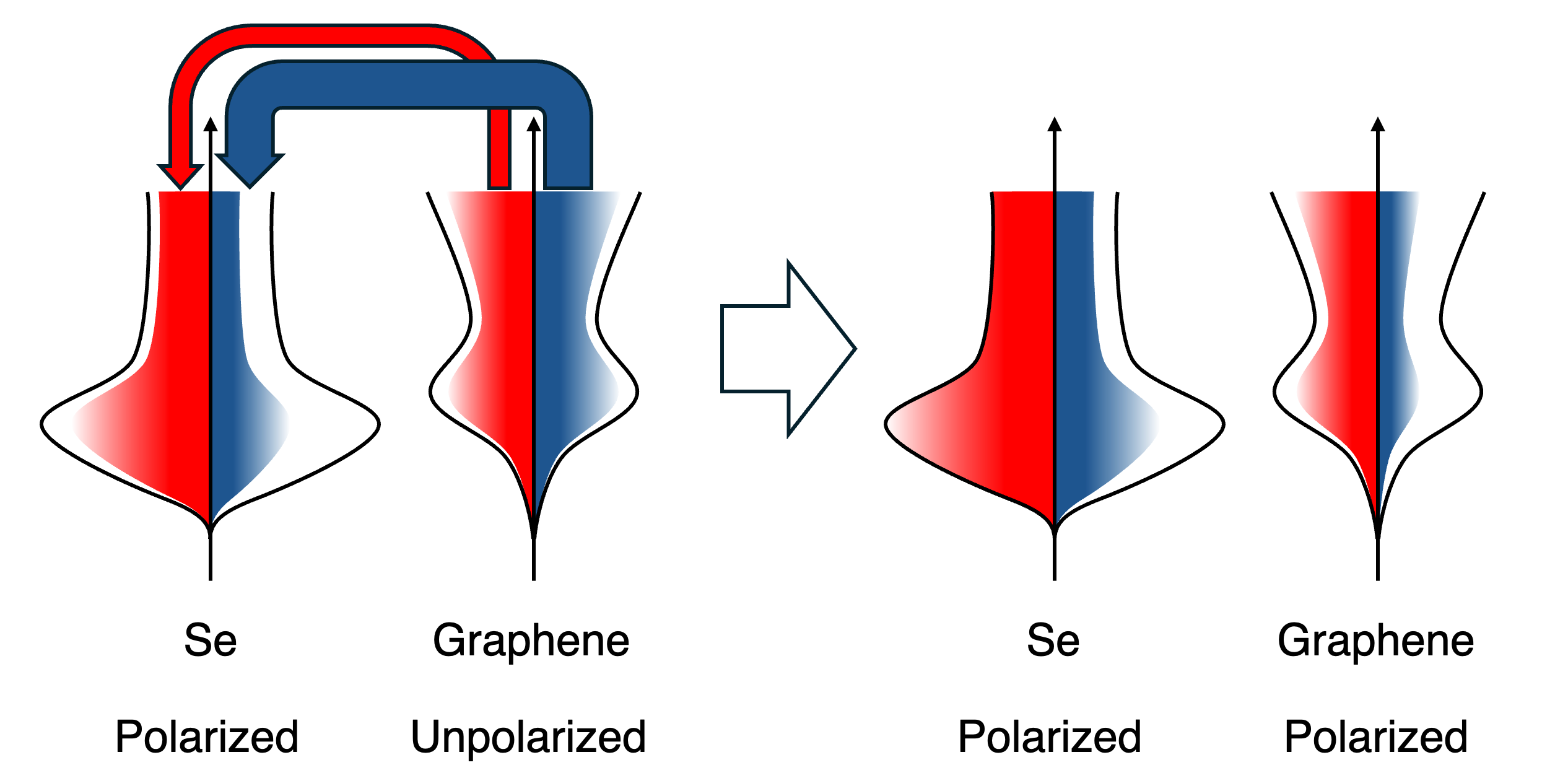}
    \caption{\label{fig:schematic}
    Schematic of spin-selective electron migration between the Se layer at $z = 1.7$~{\AA} and the graphene layer.
    }
\end{figure}

Spin-polarized carriers are generated in WSe$_2$, and the electron current from graphene to WSe$_2$ is subject to Pauli blocking, which selectively inhibits one spin channel. 
As a result, migration preferentially occurs in the opposite spin channel, which leads to an accumulation of blocked spins in the graphene layer.  
This spin-selective migration induces a correlated spin distribution between the Se layer at $z = 1.7$~{\AA} and the graphene layer (Fig.~\ref{fig:schematic}). 
The occurrence of Pauli blocking due to excited carriers is supported by the saturation of electron transfer observed in Fig.~\ref{fig:polarization}.
The similarity in the total spin magnetization between the HB and the monolayer WSe$_2$ is consistent with photoexcitation occurring almost independently in the WSe$_2$ and graphene layers (Fig.~\ref{fig:hetero_spin_decomp}). 
The observed charge and spin transfer can be attributed to vertical (energy-axis) photoexcitation within each layer, followed by horizontal ($z$-axis) carrier transport across energy levels.
Such spin transfer has been reported in a HB of magnetic and non-magnetic 2D materials under a linearly polarized pulse, but is attributed instead to intrinsic spin splitting of the static band structure~\cite{Guo2024}.
In the present case, the spin-selective carrier filtering at the interface is entirely driven by dynamical effects in the non-magnetic system.

\begin{table}
\centering
\begin{tabular}{l|ccc}
 {} & $T=10$  & $T=20$  & $T=30$  \\ \hline
 $I=10^9$  & -1.2 & 0.42 & 0.24 \\
 $I=10^{10}$  & -1.4 & 0.40 & 0.19 \\
 $I=10^{11}$  & 0.66 & 0.15 & 0.11 \\
 $I=10^{12}$  & 0.023 & 0.10 & 0.10
\end{tabular}
\caption{\label{tb}
List of the spin transfer efficiency with the pulse duration $T$ (fs) and laser intensity $I$ (W/cm$^2$).
}
\end{table}

Table~\ref{tb} lists the spin transfer efficiencies (ratio of the spin magnetization around the graphene layer to that around the WSe$_2$ layer) for various combinations of pulse duration $T$ and laser intensity $I$, where the time average is taken over the interval $[T, T + 20\,\,\mathrm{fs}]$.
This provides a rough estimate based on the short-time average.  
In the low-intensity cases with $T = 10$~fs, negative values are observed because the pulse duration is too short to induce sufficient real excitation of the total spin magnetization.  
For $T > 10$~fs, the spin transfer efficiency remains stably positive and decreases monotonically with increasing intensity.  
This trend further supports the interpretation that spin transfer arises from Pauli blocking by excited carriers.

\section{Conclusions \label{sec:conclusion}}

In conclusion, the microscopic mechanisms behind spin injection in TMD-graphene heterostructures under circularly polarized laser irradiation were analyzed using TDDFT. 
First-principles calculations reveal that Pauli blocking between WSe$_2$ and graphene under laser excitation drives spin injection via opposite spin and electron transfer. The carrier transfer occurs on an ultrafast timescale that renders relaxation effects negligible. These nonequilibrium dynamics play a crucial role in the establishment of spin polarization in graphene. These findings clarify the microscopic mechanism that underlies spin injection in non-magnetic 2D heterostructures and provide a foundation for ultrafast opto-spintronic applications.

\begin{acknowledgements}
This research was supported by a grant (No. JPMXS0118067246) under the Japanese Ministry of Education, Culture, Sports, Science and Technology (MEXT) Quantum Leap Flagship Program (Q-LEAP), and by Kakenhi grants‐in‐aid (Nos. 24K01224 and 24K17629) from the Japan Society for the Promotion of Science (JSPS). 
Calculations were performed on the Fugaku supercomputer with support from the HPCI System Research Project (Project IDs: hp240124 and hp250102), the Miyabi supercomputer, and the Wisteria supercomputer at the University of Tokyo under a Multidisciplinary Cooperative Research Program of the Center for Computational Sciences (CCS) at the University of Tsukuba.
\end{acknowledgements}

%


\end{document}